\documentstyle[prb,aps,amsmath,psfig]{revtex}  

\def\H{{\cal H}}

\def\D{{\cal D}}

\def\G{{\cal G}}

\def\tS{{\widetilde S}}

\def\bS{{\boldsymbol S}}
\def\bs{{\boldsymbol s}}
\def\bj{{\boldsymbol j}}
\def\bdelta{{\boldsymbol\delta}}
\def\bk{{\boldsymbol k}}

\begin{document}
\wideabs{
\title{Semiclassical approach to the thermodynamics of spin chains}

\author{Alessandro Cuccoli, Valerio Tognetti and Paola Verrucchi}
\address{Dipartimento di Fisica dell'Universit\`a di Firenze
        and \\ Istituto Nazionale di Fisica della Materia (INFM),
        Largo E. Fermi~2, I-50125 Firenze, Italy}

\author{Ruggero Vaia}
\address{Istituto di Elettronica Quantistica
         del Consiglio Nazionale delle Ricerche,\\
         via Panciatichi~56/30, I-50127 Firenze, Italy}

\date{\today}
\maketitle

\begin{abstract}
  Using the PQSCHA semiclassical method, we evaluate thermodynamic
  quantities of one-dimensional Heisenberg ferro- and
  antiferromagnets.  Since the PQSCHA reduces their evaluation to
  classical-like calculations, we take advantage of Fisher's exact
  solution to get all results in an almost fully analytical way.
  Explicitly considered here are the specific heat, the correlations
  length and susceptibility. Good agreement with Monte Carlo
  simulations is found for $S>1$ antiferromagnets, showing that the
  relevance of the topological terms and of the Haldane gap is
  significant only for the lowest spin values and temperatures.
\end{abstract}
}


Several applications to condensed matter systems have demonstrated the
usefulness of the improved~\cite{GT85prl,FeynmanK86} effective potential
approach~\cite{FeynmanH65}; its generalized version for non standard
Hamiltonian, the so called {\em pure-quantum self-consistent harmonic
approximation} (PQSCHA)~\cite{CTVV92ham,CGTVV95}, has also been
successfully applied to different spin systems.

We consider here the one-dimensional isotropic Heisenberg model
Hamiltonian,
\begin{equation}
 \H = \pm \frac J2 \sum_{\bj\,\bdelta} \bS_\bj{\cdot}\bS_{\bj+\bdelta} ~,
 \label{heisham}
\end{equation}
with the exchange interaction $J>0$ restricted to nearest-neighbour sites
of a simple cubic $d$-dimensional lattice; the sign $-$ refers to the
ferromagnet (FM), $+$ to the antiferromagnet (AFM). The thermodynamic
quantities of this model were successfully calculated for
two-~\cite{CTVV97prl,CTVV97prb} and three-dimensional~\cite{CMV99}
magnets, both of which are characterized by a ground state that can be
obtained perturbatively starting from the classical-like minimum-energy
configuration.

In one dimension the situation is largely different and the ferro- and
antiferromagnets, at variance with the classical case where both models
are mapped onto the classical non linear Schr\"odinger equation, behave in
a markedly different way. In fact, ferromagnets do not present any
peculiarity, being their (ordered) ground state, as in higher dimension,
the quantum counterpart of the classical minimum-energy configuration. The
relevant excitations, both linear and non-linear, which contributes to the
thermodynamic properties and destroy the long-range order at any finite
temperature, are basically the same in the quantum and in the classical
case. The absence of long-range order at any $T\neq0$~\cite{MerminW66}
implies that the linear excitations spectrum should be considered only up to
wavelengths of the order of the correlation length $\xi(T)$.

Quantum effects have a much more apparent impact on the qualitative
behaviour of antiferromagnets. Its ground state cannot be obtained
perturbatively from the N\'eel configuration, and the relevant quantum
excitations are completely different from the classical ones. Moreover, as
firstly suggested by Haldane~\cite{Haldane83a,Haldane83b}, integer and
half-integer spin chains display a qualitatively different low-temperature
behaviour, which arises from a topological term in the path-integral
description of spin systems. For half-integer spins, the interference term
leads to gapless excitations, while the {\it Haldane gap} appears for
integer $S$. However, these effects should rapidly disappear for
increasing spin value, as the interference becomes less destructive and
the gap vanishes exponentially.

Hence, in one dimensional magnets should also exist regimes where
semiclassical methods can be sensibly applied, i.e., when the spin
length and/or the temperature increase and the classical behaviour is
approached. In such regimes, the PQSCHA is a good tool for calculating
thermodynamic quantities and should be very competitive in comparison with
other semiclassical ones, as, for instance, the theory based on the use of
real-space coherent states that has been recently introduced~\cite{KladkoFG99}.
Indeed, by the PQSCHA most calculations can be performed analytically, and
with a full quantum inclusion of the linear excitations in wave-vector
space.

The final outcome of the application of the PQSCHA to Heisenberg models is
that the free energy of the quantum model described by Eq.~(\ref{heisham})
is given by the the free energy of the {\it effective classical}
Heisenberg Hamiltonian,
\begin{equation}
 \H_{\rm{eff}}=\pm \frac 12 J\tS^2\,\theta^4(t)
 \sum_{\bj\,\bdelta} \bs_\bj{\cdot}\bs_{\bj+\bdelta}+NJ\tS^2\G(t) ~,
 \label{effham}
\end{equation}
whose thermodynamic properties are exactly known after the work by
Fisher~\cite{Fisher64}. In Eq.~(\ref{effham}) ${\bs_\bj}$ are unit vectors,
$\tS=S+1/2$ plays the role of the `classical' spin length and $J\tS^2$
that of the overall energy scale: we hence define $t=k_BT/J\tS^2$ as the
reduced temperature. The temperature- and spin-dependent parameters
$\theta(t)$ and $\G(t)$ account for the effects of the {\em pure-quantum}
fluctuations. Their explicit expressions are:
\begin{eqnarray}
 \theta^2(t)&=& 1-\frac\D 2  ~,
\\
 \G(t)&=& \frac tN \sum_k
 \ln\frac{\sinh f_k}{\theta^2 f_k}-\frac z2\kappa^2(t)\,\D ~.
\end{eqnarray}
It is worthwhile pointing out that the first term of $\G(t)$ restores
the quantum free energy of the linear excitations~\cite{CGTVV95}. The
renormalization coefficient $\D(t)$ reads
\begin{equation}
 \D={1\over N\tS}\,\sum_\bk \Big(\coth f_\bk{-}\frac1{f_\bk}\Big)\times\!
 \left\{
 \begin{array}{ll}
  \sqrt{1-\gamma_\bk^2} ~~ &\mbox{(AFM)} \\
  (1-\gamma_\bk)           &\mbox{(FM)}
 \end{array} \right. ,
 \label{e.D}
\end{equation}
and represents indeed the pure-quantum nearest-neigh\-bour transverse
spin fluctuations in self-consistent Gaussian approximation. This
essential renormalization coefficient of the PQSCHA for Heisenberg
models, as well as the connected quantities $\theta^2(t)$ and $\G(t)$,
are global parameters, i.e., they take into account the quantum
effects only on average, so that the details of the excitation
spectrum are smeared out.  Furthermore,
$f_\bk=(\hbar\omega_\bk)/(2k_BT)=\widetilde\omega_\bk/2\tS{t}$, where
\begin{equation}
\widetilde\omega_\bk= \left\{
 \begin{array}{ll}
  z\kappa^2\,\sqrt{1-\gamma_\bk^2} ~~ &\mbox{(AFM)} \\
  z\kappa^2\,(1-\gamma_\bk)           &\mbox{(FM)}
 \end{array} \right. ,
\end{equation}
are the dimensionless spin-wave frequencies, whose renormalization factor
$\kappa^2(t)={1\over2}\Big(\theta^2+\sqrt{\theta^4-{4}t\varepsilon/z}\Big)$
is calculated by taking into account only the thermal fluctuations with
$|\bk|>\pi/\xi(t)$, so that $\varepsilon=1-1/\xi$ for the one-dimensional
system. We point out that the contributions to the pure-quantum
renormalization coefficient~(\ref{e.D}) are weighted by the Langevin
function $\coth{f}_\bk-f_\bk^{-1}$, so that the major role is played by
the high-frequency (short wavelength) excitations, which are just those that
survive in spite of the absence of long-range order. The PQSCHA in the
low-coupling approximation~\cite{CTVV92ham,CTVV97prl,CTVV97prb} employed
to derive Eq.~(\ref{effham}) neglects contributions of order $\D^2$ so
that $\D$ must be small compared to one.

Using the exact results for the classical one-dimen\-sional Heisenberg
model of Ref.~\onlinecite{Fisher64}, we can thus easily write an analytical
expression for the free energy per spin of the quantum spin chain within
the PQSCHA:
\begin{equation}
 {f\over J\tS^2}= \G(t,S)-t\ln\left[
 \sinh\left({\theta^4(t,S)\over t}\right)\cdot{t\over\theta^4(t,S)}
 \right]~. \label{freeen}
\end{equation}
The other macroscopic thermodynamic quantities, e.g. internal energy and
specific heat, can be easily obtained from this equation by (numerical)
derivation, taking care of the $t$-dependence of $\G$ and $\theta$, which
prevents us from using directly the expressions of Ref.~\onlinecite{Fisher64} in
evaluating such quantities. One can also obtain the spin correlation
function and the susceptibility by means of the formulas reported in
Ref.~\onlinecite{CTVV97prb}, where it is shown also how the correlation length
$\xi(t)$ is simply related to its classical counterpart only by the change
of the temperature scale involved in the renormalization of the exchange
constant,
\begin{equation}
 \xi(t)=\xi_{\rm{cl}}\big(t/\theta^4(t)\big) ~.
\end{equation}
This formula is of remarkable simplicity, especially when one notices that
$\xi_{\rm{cl}}(t)$ has a very simple analytical
expression~\cite{Fisher64},
\begin{equation}
 \xi_{\rm{cl}}(t)=\big[-\ln(\coth t^{-1}-t)\big]^{-1} ~.
\end{equation}
\medskip

\begin{figure}[hbt]
\centerline{\psfig{bbllx=0pt,bblly=0pt,bburx=612pt,bbury=792pt,%
  figure=fmd.eps,width=88mm,angle=270}}
 \caption{Pure-quantum renormalization coefficient ${\cal{D}}(t,S)$ vs.
temperature $t=T/J\tS^2$, for the Heisenberg FM chain. From the uppermost
curve, the spin values are $S=1/2$, $S=1$ and $S=3/2$.
 \label{f.Dfm} }
\end{figure}

\begin{figure}[hbt]
\centerline{\psfig{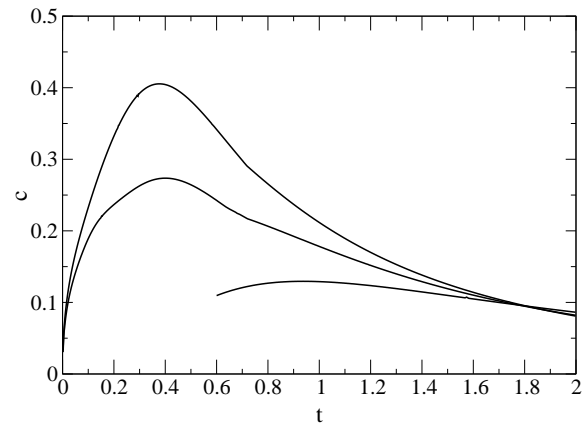}}
 \caption{Specific heat $c(t)$ of the Heisenberg FM chain for
   $S=1/2$, $S=1$ and $S=3/2$, respectively, starting from the
   lower-most curve.
 \label{f.Cfm} }
\end{figure}

In principle, the PQSCHA approach should work well for the one-dimensional
Heisenberg FM, because: ~{\it i}) their ground state is ordered and is an
eigenstate of $S^z_{\rm{tot}}=\sum_i S_i^z$; ~{\it ii}) the absence of
long-range order at any finite temperature is due to nonlinear
excitations, which can be well treated in semiclassical ($1/S$-expansion)
approximation; ~{\it iii}) the pure-quantum fluctuations related to the
linear excitations are accounted for through the coefficient $\D(t)$.
Therefore, the only limitation we should care of for the FM is a possible
too high value of $\D$: we shall assume that the condition
$\D\lesssim{0.5}$ must be satisfied to get reliable results. As shown in
Fig.~\ref{f.Dfm} this occurs at any temperature for $S\geq{3/2}$. As for
lower values of $S$ it is apparent that for $S=1$ the approach can be used
in a wide range of temperatures starting from $t\gtrsim{0.25}$; on the
other hand, the strongest quantum case, $S=1/2$, cannot be well described
except for very high temperatures $t\gtrsim{1}$.

Unfortunately, to the best of our knowledge, reference data on
one-dimensional Heisenberg ferromagnets are not available for intermediate
spin values, and we found only rather old ones for
$S=1/2$\cite{BonnerF64,CullenL83}. However it could be appealing to derive
some thermodynamic quantities by means of the effective Hamiltonian
(\ref{effham}) and Eq.~(\ref{freeen}). The specific heat of the FM spin
chain is shown in Fig.~\ref{f.Cfm} for spin values $S=1/2$, $S=1$ and
$S=3/2$. Comparing with Fig.~\ref{f.Dfm} we see why the curve for $S=1/2$
is truncated at $t\simeq{1}$: for higher temperature the behaviour of
$C(t)$ is in agreement with the available numerical
data\cite{BonnerF64,CullenL83} .

\begin{figure}[hbt]
\centerline{\psfig{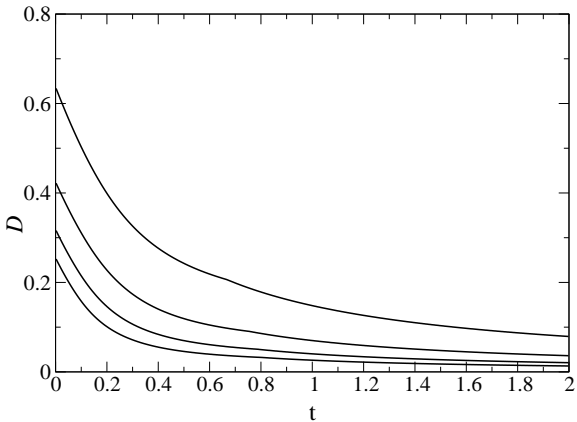}}
 \caption{Pure-quantum renormalization coefficient $\D(t,S)$ vs.
  temperature $t=T/J\tS^2$, for the Heisenberg AFM chain.
  From the uppermost curve the spin values are $S=1/2$, $1$,
  $3/2$, and $2$.
 \label{f.Dafm} }
\end{figure}

Turning to antiferromagnets, the temperature behaviour of $\D(t)$ is
shown in Fig.~\ref{f.Dafm}. Using the same criterion as above, we can
deduce that the PQSCHA should work well for any temperatures for
$S\ge{1}$ while its validity is confined to $t\gtrsim{0.15}$ for
$S=1/2$. However, we must recall that the afore-mentioned quantum
effects which strongly modify the nature of the ground state and lead
to the Haldane gap, cannot be accounted for by any semiclassical
approach, so that we do not expect that peculiar quantum properties at
very low temperature and low values of the spin could be reproduced
within the PQSCHA. Instead the PQSCHA is expected to give good results
when the spin becomes larger and larger approaching the classical AFM
Heisenberg model.

\begin{figure}[hbt]
\centerline{\psfig{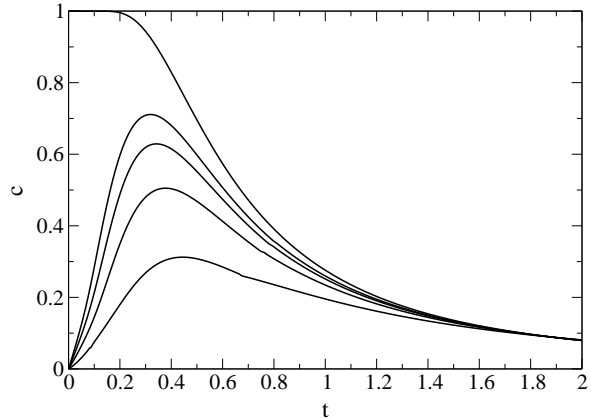}}
 \caption{Specific heat $c(t)$ of the Heisenberg AFM chain for
   $S=1/2$, $1$, $3/2$, $2$, and $\infty$ (classical limit),
   respectively, starting from the lower-most curve.
 \label{f.Cafm} }
\end{figure}

\begin{figure}[hbt]
\centerline{\psfig{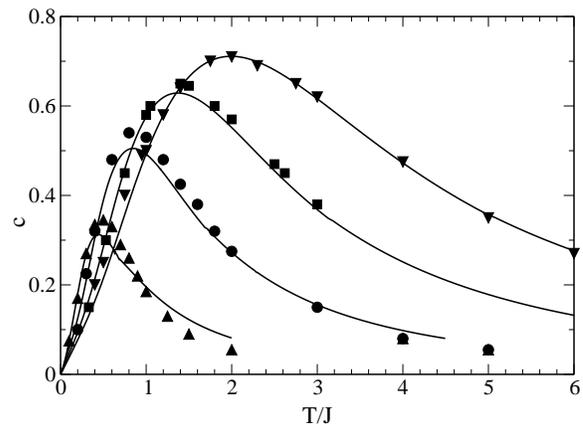}}
 \caption{Comparison of the specific heat $c(t)$ of the Heisenberg
   AFM chain, plotted  vs. $T/J=t\tS^2$, with numerical simulation data for
   $S=1/2$~\protect\cite{Xiang98}, $S=1$~\protect\cite{YamamotoM93},
   $S=3/2$~\protect\cite{Xiang98} and $S=2$~\protect\cite{Yamamoto96}.}
 \label{f.Cafmmc}
\end{figure}

The results for the specific heat plotted in Figs.~\ref{f.Cafm}
and~\ref{f.Cafmmc} seem to confirm this prediction for the thermodynamic
quantities. The comparison with the existing quantum Monte Carlo
\cite{YamamotoM93,Yamamoto96} and transfer-matrix Renormalization Group
\cite{Xiang98} data shows that the agreement improve for larger spin and
can be considered very good almost at all temperature already for $S=3/2$.

\begin{figure}[hbt]
\centerline{\psfig{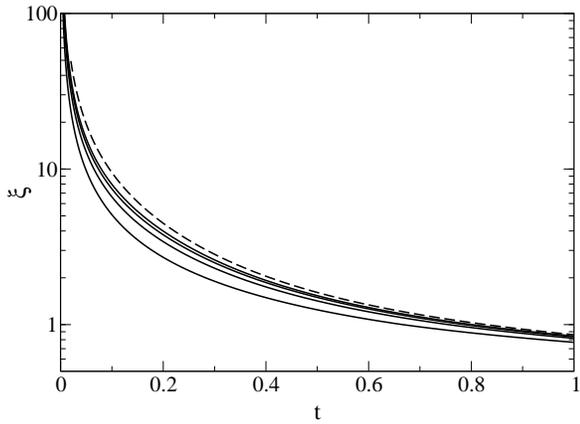}}
 \caption{Correlation length $\xi(t)$ of the AFM Heisenberg chain for
   $S=1/2$, $S=1$, $S=3/2$ and $S=2$, respectively, starting from the
   lower-most curve; the dashed line is the classical result.
 \label{f.xiafm} }
\end{figure}

\begin{figure}[hbt]
\centerline{\psfig{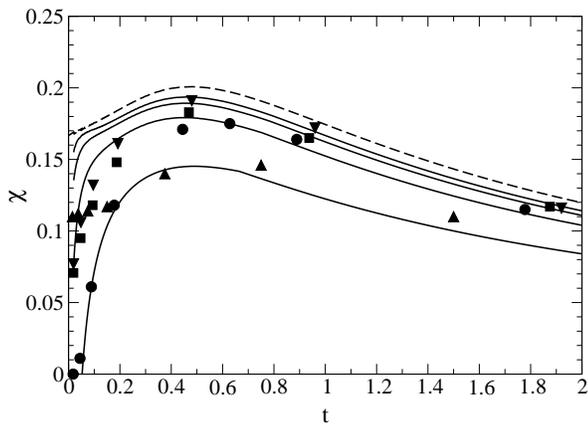}}
 \caption{Comparison of the uniform susceptibility $\chi(t)$ of the
   AFM Heisenberg chain with Quantum Monte Carlo simulation
   data~\protect\cite{KimGWB98}. Spin values are
   $S=1/2$ (line and upward triangles), $S=1$ (line and circles),
   $S=3/2$ (line and squares) and $S=2$ (line and downward triangles),
   respectively, starting from the lower-most curve; the dashed line
   is the classical result.
 \label{f.chiafm} }
\end{figure}

In Figs. \ref{f.xiafm} and \ref{f.chiafm} we show the PQSCHA results for
the correlation length $\xi(t)$ and the uniform susceptibility $\chi(t)$
of the AFM chain. The latter quantity has been shown to be particularly
sensitive to test the peculiar quantum effects related to the Haldane gap;
nevertheless, Fig. \ref{f.chiafm} clearly shows that a semiclassical
approach like the PQSCHA compares well with numerical data not only where
the classical regime is already approached, but also in the intermediate
temperature range, where the curves for low spin values already are
clearly different from the classical behaviour; only the very low
temperature behaviour can not be, as expected, reproduced.

The conclusion is that the thermodynamic quantities of the Heisenberg AFM
are not strongly affected by the consequences of the Haldane ground state
at intermediate and high temperatures. Small differences appear only at
the lowest values of the spin and from the point of view of the PQSCHA it
is difficult to ascertain if they are due to the "Haldane effects" or to
the rather large value of $\D$. On the other hand, it is curious to note
that the Haldane conjecture was theoretically proven for large $S$ --
where the role of the topological term decreases and the gap is
exponentially vanishing, giving negligible quantitative contributions to
the thermodynamics.



\end{document}